\documentstyle[prb,aps,twocolumn]{revtex}
\large
\begin{document}
\draft
\twocolumn[\hsize\textwidth\columnwidth\hsize\csname
@twocolumnfalse\endcsname

\title{The magnetic ordering in the mixed valence compound
$\beta$-Na$_{0.33}$V$_2$O$_5$}

\author{A.~N.~Vasil'ev}
\address{M.~V.~Lomonosov Moscow State University,
119899 Moscow, Russia}

\author{V.~I.~Marchenko, A.~I.~Smirnov, S.~S.~Sosin }
\address{P.~L.~Kapitza Institute for Physical Problems RAS, 117334
Moscow, Russia}

\author{H.~Yamada, Y.~Ueda}
\address{Institute for Solid State Physics, University of Tokyo, 7-22-1
Roppongi, Minato-ku, Tokyo 106, Japan}

\date{\today}
\maketitle

\begin{abstract}
\widetext
\leftskip 54.8pt
\rightskip 54.8pt

The low-temperature electron spin resonance (ESR) spectra and the
static magnetization data obtained for the stoichiometric single
crystals of $\beta$-Na$_{0.33}$V$_2$O$_5$ indicate that this
quasi-one-dimensional mixed valence compound demonstrates at
$T_N=22$~K the phase transition into the canted
antiferromagnetically ordered state. The spontaneous magnetization
of $3.4\times 10^{-3}$ $\mu_B$ per V$^{4+}$ ion was found to be
oriented along the two-fold $b$ axis of the monoclinic structure,
the vector of antiferromagnetism is aligned with the $a$ axis and
the Dzyaloshinsky vector is parallel to the $c$-axis. The
experimental data were successfully described in the frame of the
macroscopic spin dynamics and the following values for the
macroscopic parameters of the spin system were obtained: the
Dzyaloshinsky field $H_D=6$~kOe, the energy gaps of two branches
of the spin wave spectrum $\Delta_1=48$~GHz and $\Delta_2=24$~GHz.

\end{abstract}


]

\narrowtext

\section{Introduction}

The quasi-one-dimensional mixed valence compound
$\beta$-Na$_{0.33}$V$_2$O$_5$ exhibits various phase transitions
of a considerable interest.\cite{Ozerov,Schlenker,YamadaUeda}
Below 230~K the $1\times 2 \times 1$ crystal lattice
superstructure appears, indicating probably the ordering of Na
ions. At $T_c=136$~K this one-dimensional conductor exhibits a
metal-insulator phase transition of a charge ordering type and at
$T=T_N=22$~K the charge-ordered structure undergoes an
antiferromagnetic phase transition.

This behavior is to be compared with that of the
quasi-one-dimensional magnet $\alpha$-NaV$_2$O$_5$ in which the
lattice, charge and spin subsystems transform simultaneously. The
charge ordering in this related system causes the opening of the spin
gap in the spectrum of magnetic excitations and the appearance of a
nonmagnetic singlet state.  \cite{IsobeUeda,Fujii,Smirnov}

The $\beta$-Na$_{0.33}$V$_2$O$_5$ has the monoclinic structure
C2/m (see Fig.~1) with the unit cell containing two
NaV$_6$O$_{15}$ formula units. The sodium ions are located in the
tunnels formed by a V-O framework. The Na$^+$ ions occupy only one
of the two nearest-neighboring sites $A$ in the $ac$-plane. The
V-O framework consists of the three distinct kinds of double
chains directed along the $b$-axis.\cite{Wadsley,Deramond} The
V1-sites have a six-fold octahedral coordination and form a zigzag
chain of edge-sharing VO$_6$ octahedra. The V2-sites with a
similar octahedral coordination form a two-leg ladder chain of
corner-sharing VO$_6$ octahedra, and the V3-sites, having a
five-fold square pyramidal coordination form a zigzag chain of
edge-sharing VO$_5$ pyramids.

The absence of the Knight shift in NMR experiments on $^{23}$Na nuclei
\cite{Maruyama} shows that the outer $s$-shell electrons of the Na-ions
are transferred into the $d$-shells of V-ions.  Therefore, the V-ions
are in the mixed valence states V$^{4+}$ and V$^{5+}$. The V$^{4+}$
ions are in a magnetic $S=\frac{1}{2}$ state, while the V$^{5+}$  ions
with $S=0$ are nonmagnetic. Basing on interatomic distances
\cite{Goodenough} and on NMR measurements \cite{NMR2,NMR3} it was
concluded that in the high temperature phase $T>T_c$ the donated
electrons are situated at the V1 sites, with one half of these sites
being V$^{4+}$.  The recent NMR experiments on V-ions \cite{Itoh}
confirm the charge ordering nature of the transition at $T=T_c$ and
reveal that the number of inequivalent V positions below $T_c$ is
increased. Two possible models of the charge ordering (zigzag and
linear chains of V$^{4+}$ ions) are proposed but the exact crystal
structure in a low temperature phase is still unknown.

Hypothetical models of the magnetic structure with the localized
electrons suggest  that only one-sixth part of the V ions has the
magnetic moments and the magnetic subsystem of
$\beta$-Na$_{0.33}$V$_2$O$_5$ appears to be strongly diluted.
Nevertheless, the transition to the long-range magnetically ordered
state occurs at an appreciably high temperature $T_N=22$~K. The aim of
the present work was to study in details the magnetic ordering in
$\beta$-Na$_{0.33}$V$_2$O$_5$ by means of magnetic resonance and static
magnetization measurements.

\section{Experimental}

\subsection{Samples and the experimental techniques}

The single crystals of a stoichiometric
$\beta$-Na$_{0.33}$V$_2$O$_5$ were grown by a self flux technique
using NaV$_3$O$_8$ as a flux. The crystals of a typical size
$5\times 0.5\times 0.2$~mm$^3$ were obtained by melting  the
mixture of one part of Na$_{0.33}$V$_2$O$_5$ powder and thirty
parts (weight ratio) of NaV$_3$O$_8$ powder. The melting at 740
$^{\circ}$C and cooling down to 600 $^{\circ}$C with a rate of 0.5
$^{\circ}$C/h were performed in a vacuum. The flux was removed by
diluted hydrochloric acid. The longest dimension of the
rectangular sample was aligned with the $b$ axis, the middle
dimension was aligned with the $c$ axis, and the shortest
dimension, denoted below as $a^{\prime}$ was perpendicular to the
$(bc)$-plane. The crystallographic $a$ axis is directed at an
angle of 18$^{\circ}$ with respect to $a^{\prime}$ axis in the
$a^{\prime}c$ plane. The static magnetization measurements were
done by the Quantum Design SQUID magnetometer. The magnetic
resonance measurements in the frequency range 18-80 GHz were
performed by the transmission type microwave spectrometer. The
sample was put into the rectangular resonator having a set of
eigen-frequencies in this range.

The temperature dependencies of the magnetization of
$\beta$-Na$_{0.33}$V$_2$O$_5$ sample measured along the
$a^\prime$, $b$ and $c$ axes at the magnetic field $H=10$ kOe are
shown in Fig.~2. The change of the slope on these curves at
$T_c$=136~K marks the metal-insulator transition associated with
the charge redistribution between the vanadium sites. The sharp
increase in magnetization is observed at $T=T_N$.

The field dependencies of magnetization measured along the
$a^{\prime}$, $b$ and $c$ axes at $T=5$~K are shown in Fig.~3. The
asymptotically linear behavior of the $M(H)$ dependencies indicate
the existence of the antiferromagnetic ordering. The residual
magnetization $M(H=0)$ for ${\bf H}\parallel b$ (as well as the
anisotropic anomaly on $M(T)$ curves at $T=T_N$) are the evidence
of a canting of the antiferromagnetic sublattices.
The nonlinear part of the $M(H)$ curve at ${\bf H}\parallel
a^\prime$ should result from a spin-reorientation process. The
extrapolation of $M(H)$ to $H=0$ at ${\bf H}\parallel b$ gives a
finite value of the spontaneous magnetization with a weak
ferromagnetic moment of about $3\times 10^{-3} \mu_B$ per V$^{4+}$
ion.

The ESR absorption lines of $\beta$-Na$_{0.33}$V$_2$O$_5$ obtained
as the field dependencies of the signal transmitted through the
resonator with the sample at various temperatures are shown in
Fig.~4. In the paramagnetic state (at $T>T _N$) the ESR line is
observed at a slightly larger field than the DPPH reference
($g=2.0$). Below $T_N$ the ESR line rapidly shifts to lower
magnetic fields.

The evolution of the ESR absorption with the frequency at
$T=4.2$~K is illustrated in Fig.~5. Up to four resonant lines
marked by letters A, B, C, D were present in the absorption at
different frequencies and sample orientations. The C and D lines
disappear above 10~K and could not be ascribed to the ordered
state, while the A and B lines exist in the whole temperature
range below $T_N$ and thus, should be of an antiferromagnetic
type.

The magnetic resonance data at $T=4.2$~K are summarized in
Figs.~6a,b,c, in which the frequencies of the resonant lines are
plotted {\it vs}  magnetic field oriented along the
$a^{\prime}~,b$ and $c$ axes respectively.
The spectrum of the magnetic resonance in the ordered phase
appears to be highly anisotropic. It consists of two branches with
the gaps of 48 GHz and 24 GHz.
The falling branch of the magnetic resonance spectrum at ${\bf H}
\parallel a^{\prime}$ and the drop-to-zero of its resonance frequency
at $H=6$ kOe clearly indicate the spin-reorientation process ended
by a phase transition of the second kind. The characteristic
feature of the field dependencies of the resonance frequencies at
${\bf H}\parallel a^{\prime}$ and $b$ is the nonzero slope of both
branches. The smaller of these slopes reveals the existence of
canting of sublattices.

\section {Discussion}

The general form of the antiferromagnetic resonance spectra and
the magnetization curves prove the appearance of the canted
antiferromagnetic state in $\beta$-Na$_{0.33}$V$_2$O$_5$ below
$T<T_N$. The detailed analysis given below will allow us to obtain
its macroscopic parameters.

Using the phenomenological approach \cite{AndreevMarchenko} we
calculated the field dependencies of the static magnetization and
the resonance spectra of a canted antiferromagnet with two axes of
anisotropy for the various orientations of the magnetic field.
According to this approach the antiferromagnetic structure is
considered to be collinear in the exchange approximation, and the
effects resulting from relativistic interactions (the sublattices
canting and the anisotropy) are taken into account as
perturbations. At low temperature the Lagrange function of a
mole of such an antiferromagnet may be represented in the form:

\begin{equation}
{\cal L}= \frac{\chi_{\perp}}{2 \gamma^2} ( \dot{{\bf l}} -
\gamma[{\bf lH}] ) ^2 - \frac{\chi_{\perp}}{ \gamma}({\bf d},\dot
{{\bf l}}-\gamma[{\bf lH}]) - U_a, \label{Lagr}
\end{equation}

\noindent where ${\bf l}$ is the unit vector of the order
parameter, $\chi_{\perp}$ is the antiferromagnetic susceptibility,
$\gamma$ is the magnetomechanical ratio (suggested to be equal to
that of a magnetic V ion $\gamma=g\mu_B/h$), ${\bf d}$ is the
Dzyaloshinsky vector giving rise to the spontaneous magnetization
${\bf M}_{sp}=\chi_{\perp}[{\bf dl}]$, $U_a$ is the potential
energy of the relativistic anisotropy taken as the quadratic form
in the vector ${\bf l}$ components:

\begin{equation}
U_a=\frac{\beta_1}{2}({\bf xl})^2 + \frac{\beta_2}{2}({\bf yl})^2.
\label{anis}
\end{equation}

Here $\beta_1, \beta_2 $ are positive constants of anisotropy,
unit vectors ${\bf x}$ and ${\bf y}$ determine two orthogonal
directions in the crystal. One of these two directions
(corresponding to the largest coefficient $\beta_i$) should be the
hard direction for the spin structure, the other  one should be
the middle direction, the third orthogonal unit vector ${\bf z}$
will be an easy axis (the direction of the vector ${\bf l}$ at
${\bf H=0}$). The orientation of vectors ${\bf x}$, ${\bf y}$,
${\bf z}$ and ${\bf d}$ with respect to the crystal axes $a$, $b$
and $c$ is {\it a priori} not known, except for the condition that
one of the vectors ${\bf x}$, ${\bf y}$ or ${\bf z}$ should be
parallel to the two-fold axis $b$.

The total static magnetization of this system is:

\begin{equation}
{\bf M} =\frac{\partial {\cal L}}{\partial {\bf H}} =
\chi_{\perp}({\bf H} - {\bf l}({\bf lH}) + [{\bf dl}]).
\label{Moment}
\end{equation}

The equilibrium value of the vector ${\bf l}$ is determined by
minimizing the potential energy of the system ${\cal P} = - {\cal
L}(\dot{{\bf l}}=0)$. The magnetic moment $M(H)$ measured in
the experiment is the projection of ${\bf M}$ onto ${\bf H}$.

The crystal symmetry requires the spontaneous moment ${\bf
M}_{sp}$ to be either parallel or perpendicular to the two-fold
$b$ axis. The static magnetization measurements confirm the first
orientation. At the same time the extrapolation of $M(H)$ at ${\bf
H}\parallel c$ to $H=0$ gives no net magnetic moment. In
accordance with formula (\ref{Moment}) it means that the vector
product $[{\bf dl}]$ should be perpendicular to ${\bf H}$ in high
fields. The only possibility to satisfy this condition is to align
vector ${\bf d}$ with the $c$ axis.

The easy direction ${\bf z}$ of a canted antiferromagnet should be
perpendicular to ${\bf M}_{sp}$ and hence, lie in the $ac$ plane.
Thus, either ${\bf x}$ or ${\bf y}$ should be parallel to $b$
axis. Taking this orientation for vector ${\bf y}$, we have also
the vector ${\bf x}$ in the $ac$ plane. We shall denote the angle
between the vectors ${\bf d}$ and ${\bf x}$ as $\alpha$ (see the
inset to Fig.~3). This angle is not determined by the monoclinic
symmetry and in principle may be arbitrary.

To describe the dynamical properties of the system one should find
the variation of the Lagrange function (\ref{Lagr}) by ${\bf l}$:

\begin{equation}
\delta {\cal L} = \left ( \frac{\partial {\cal L}}{\partial {\bf
l}}-\frac{d}{dt} \frac{\partial {\cal L}}{\partial\dot{\bf l}}
\right ){\bf\delta l}. \label{var}
\end{equation}

\noindent
Taking ${\bf\delta l}=[{\bf \delta  \theta l}]$, where $\delta\theta$
is the vector of a small rotation, we obtain the following
system of nonlinear equations:

\begin{eqnarray}
[{\bf l},-\ddot{\bf l}+2\gamma [\dot{\bf l}{\bf H}] +
\gamma^2[{\bf H}[{\bf lH}]]+ \gamma^2[{\bf Hd}] - \Delta_1^2 {\bf
x}({\bf xl})-
\nonumber \\
  - \Delta_2^2 {\bf y}({\bf yl})] = 0, \label{equation}
\end{eqnarray}

\noindent where
$\Delta_{1,2}^2=\gamma^2\frac{\beta_{1,2}}{\chi_{\perp}}$ are two
independent phenomenological parameters corresponding to two
energy gaps of the spectrum. The linearized equations obtained
from (\ref{equation}) by expanding ${\bf l}$ in the vicinity of
equilibrium give the resonance spectrum of the system.

When the vectors ${\bf d}$ and ${\bf z}$ are mutually
perpendicular i.e. at $\alpha = 0$ the resonant frequencies may be
calculated analytically for three basic orientations of the
magnetic field.

\begin{enumerate}

\item ${\bf H}\parallel {\bf z}$: in this case the spin reorientation occurs
under magnetic field. It is the continuous rotation of the vector
${\bf l}$ in the plane containing easy and middle axes. Two
intervals of the magnetic field should be considered separately.

\begin{itemize}
\item[a)] $H < H_c$: the equilibrium angle $\psi$ between ${\bf
l}$ and ${\bf z}$ is determined by the condition

\begin{equation}
\sin{\psi} = \frac{H_DH}{\Delta_j^2/\gamma^2-H^2}, \label{sinus}
\end{equation}

\noindent where $H_D=|{\bf d}|$, $\Delta_j$ is a smaller of the
constants $\Delta_{1,2}$. The rotation terminates at $H=H_c$
determined by the relation $\sin{\psi}=1$, when ${\bf l}$ becomes
perpendicular to ${\bf H}$.

The resonance frequencies $\omega_1$ and $\omega_2$ are the roots
of the biquadratic equation:

\begin{equation}
(\omega^2+ A)(\omega^2+ B)-4\gamma^2H^2 \cos^2{\psi}\omega^2=0,
\label{hz1}
\end{equation}

\noindent where $A = \gamma^2 (H^2 \cos^2{\psi}+H_DH\sin{\psi})
-\Delta_i^2+\Delta_j^2\sin^2{\psi}$, \\ $B = (\gamma^2 H^2 -
\Delta_j^2)\cos{2\psi}+\gamma^2 H_DH\sin{\psi}$, $\Delta_i$ is the
largest of $\Delta_{1,2}$.

\item[b)] $H>H_c$:

\begin{eqnarray}
\omega_1^2 & = & \Delta_i^2-\Delta_j^2+\gamma^2HH_D \nonumber \\
\omega_2^2 & = & \gamma^2 H(H+H_D)-\Delta_j^2. \label{hz2}
\end{eqnarray}

\end{itemize}

\item ${\bf H}\parallel {\bf x}$:

\begin{eqnarray}
\omega_1^2 & = & \gamma^2H^2+\Delta_1^2 \nonumber \\
\omega_2^2 & = & \Delta_2^2.
\label{hx}
\end{eqnarray}

\item ${\bf H} \parallel {\bf y}$:

\begin{eqnarray}
\omega_1^2 & = & \Delta_1^2 \pm \gamma^2HH_D \nonumber \\
\omega_2^2 & = & \Delta_2^2 \pm \gamma^2 H(H+H_D). \label{hy}
\end{eqnarray}

\noindent The signs $\pm$ correspond to the domains of positive
and negative directions of the spontaneous magnetization with
respect to the magnetic field. On increasing the field the
negative domain disappears and the corresponding resonance line
disappears too according to our observations.
\end{enumerate}

The resonance frequencies calculated by formulae
(\ref{hz1}-\ref{hy}) are represented in Figs.~6a,b,c by dashed
lines. For an arbitrary value of $\alpha$ ($\alpha \neq 0$) the
equilibrium orientation of ${\bf l}$ in magnetic field was found
numerically (for three orientations ${\bf H}\parallel
a^{\prime},~b,~c$) and then substituted to the expression for the
magnetization and to the appropriate linearized equations of
motion.

Finding the antiferromagnetic susceptibility $\chi_{\perp}\approx
3.2\times 10^{-3}$ emu/mol V$^{4+}$ from the slope of $M(H)$ curve
at ${\bf H}\parallel c$ and varying the other four parameters
$\alpha$, $H_D$, $\Delta_1$ and $\Delta_2$ one can fit the
magnetization curves and the resonance spectra for all three
orientations.
The best fit to experimental data was obtained for $\alpha
=18^{\circ}$, $\Delta_1=48$ GHz ($\beta_1=1.1\times 10^{-2}$ K per
V$^{4+}$ ion), $\Delta_2=24$ GHz ($\beta_2=2.7\times 10^{-3}$ K
per V$^{4+}$ ion) and $H_D=6$~kOe.
The results of this fitting are shown  in Fig.3 and Fig.6 by solid
lines. The determined value of $\alpha=18^{\circ}$ means that the easy
direction of the spin system ${\bf z}$ is near the crystallographic
axis $a$. It should be noted that the resonance spectrum at $H\parallel
a^{\prime}$ is very sensitive to the value of $\alpha$ due to the
effect of dynamic repulsion between two antiferromagnetic resonance
branches.\cite{BorovicProzorova} The intersection of the two
branches may be observed at the exact orientation of the field
along the $z$ or $y$ directions (it was observed at ${\bf
H}\parallel {\bf y}$, see Fig. 6b). The "repulsion" is clearly
seen in Fig.~6a at ${\bf H}\parallel a^{\prime}$ when the magnetic
field appeared to be tilted with respect to $z$. The relation
$\Delta_1 > \Delta_2$ indicates that ${\bf x}$ is the hard axis
and ${\bf y}$ is the middle axis of the ordered spin structure.
The mismatch between the observed and calculated resonance
frequencies for the first branch in low fields at ${\bf
H}\parallel b$ remains unclear.

The magnetization is also described satisfactorily for ${\bf
H}\parallel c$ and for ${\bf H}\parallel b$ (except for the small
low field region associated with the poling). The agreement for
the third orientation ${\bf H}\parallel a^{\prime}$ appears to be
only qualitative, but still covers the most remarkable feature --
the low field nonlinear part (mentioned in the Section 2)
appearing due to the process of the continuous spin reorientation.

The above approach to the description of the uniform magnetic
properties of a spin system is self-consistent and does not
require any suggestions except that the exchange structure of the
system is not strongly distorted by relativistic interactions. To
evaluate the exchange integral from
our data we will for simplicity neglect for the effect of zero point
fluctuations \cite{Zhitomirsky}  on the susceptibility of
quasi-one-dimensional system.
In the molecular field approximation one can obtain the
intrachain exchange integral
$J = (g\mu_B)^2 N_A/4k_B\chi_{\perp}\simeq 120$~K
which corresponds to the molecular field $H_e=1700$~kOe. The
approximate evaluation for the Neel temperature of a
quasi-one-dimensional antiferromagnet $T_N \sim
\sqrt{JJ_{\perp}}$ gives the value for the interchain
exchange interaction $J_{\perp} \sim 1$~K.

For anisotropy fields $H_{A1}$ and $H_{A2}$ corresponding to the
values $\beta_{1,2}$ we have the following estimations: $H_{A1}=
340$ Oe, $H_{A2} = 85$ Oe. From the value of Dzyaloshinsky field
$H_D=6$ kOe we obtain that the spontaneous magnetization $M_{sp}=
\chi_{\perp}H_D = 3.4 \times 10^{-3} \mu_B$ per V$^{4+}$ ion which
is in qualitative agreement with the data of
Ref.~\CITE{Schlenker}.

\section{Conclusions}
The magnetic resonance spectra and the magnetization curves obtained at
low temperatures in the mixed valence charge ordered
quasi-one-dimensional compound $\beta$-Na$_{0.33}$V$_2$O$_5$ confirm
the transition into the canted antiferromagnetic state.  The
easy-direction of the spin system and the vector of the spontaneous
magnetization are found to be aligned with the $a$ axis and $b$ axis
correspondingly. The Dzyaloshinsky vector was found to be directed
along $c$ axis. The hard axis of the spin structure lies in the $ac$
plane at an angle of 18$^{\circ}$ with respect to $c$ axis, direction
$b$ is the middle axis of the spin anisotropy.  The values of the intra
and interchain exchange integrals as well as the anisotropy fields and
the Dzyaloshinsky field are obtained.

\section {Acknowledgments}

Authors thank H.Nojiri for cooperation in static magnetization
measurements, L.A. Prozorova and V.D. Buchelnikov for valuable
discussions. This work was supported by the Russian Fund for Basic
Researches grants 00-02-17317 and 99-02-17828, INTAS grant
99-0155, NWO grant 047-008-012 and CRDF grant RP1-2097.

{\bf Figure captions}

\noindent Fig. 1. The crystal structure of
$\beta$-Na$_{0.33}$V$_2$O$_5$.

\noindent Fig. 2. The temperature dependencies of the
susceptibility $M/H$ for three principal orientations of the
magnetic field.

\noindent Fig.3. The magnetization curves at $T = 5$~K\\ Inset:
The orientation of the Dzyaloshinsky vector ${\bf d}$,  hard,
middle and easy axes (see the text) with respect to
crystallographic axes.

\noindent Fig. 4.  The temperature evolution of the ESR absorption
line measured at the frequency $f=24$~GHz (${\bf H} \parallel b$).

\noindent Fig. 5. The ESR lines at different frequencies at
$T=4.2$~K (${\bf H}\parallel b$).

\noindent Fig. 6. The ESR spectra for three orientations of the
magnetic field at $T=4.2$~K. Symbols are the experimental data
(closed ones on the panel b represent the low intensity absorption
ascribed to "negative" domains). The dashed lines are drawn by
formulae (\ref{hz1}-\ref{hy}) with $\Delta_1=48$ GHz,
$\Delta_2=24$ GHz and $H_D=6$ kOe; solid lines are the result of
theoretical calculations (see text) for "positive" domains,
dashed-dotted lines are those for "negative" domains.

\end{document}